\documentclass[journal=ancac3,manuscript=article]{achemso}

\usepackage{chemformula} 
\usepackage[T1]{fontenc} 

\usepackage{amsmath}
\usepackage{xcolor}

\author{Rapha\"{e}l Sarfati}
\author{Daniel K. Schwartz}
\email{daniel.schwartz@colorado.edu}
\affiliation[University of Colorado Boulder]
{Department of Chemical and Biological Engineering, University of Colorado Boulder, Boulder, Colorado 80309, USA}

\title[]{Temporally Anticorrelated Subdiffusion in Water Nanofilms on Silica Suggests Near-Surface Viscoelasticity}


\begin{document}

\begin{tocentry}

\includegraphics[width=0.9\textwidth]{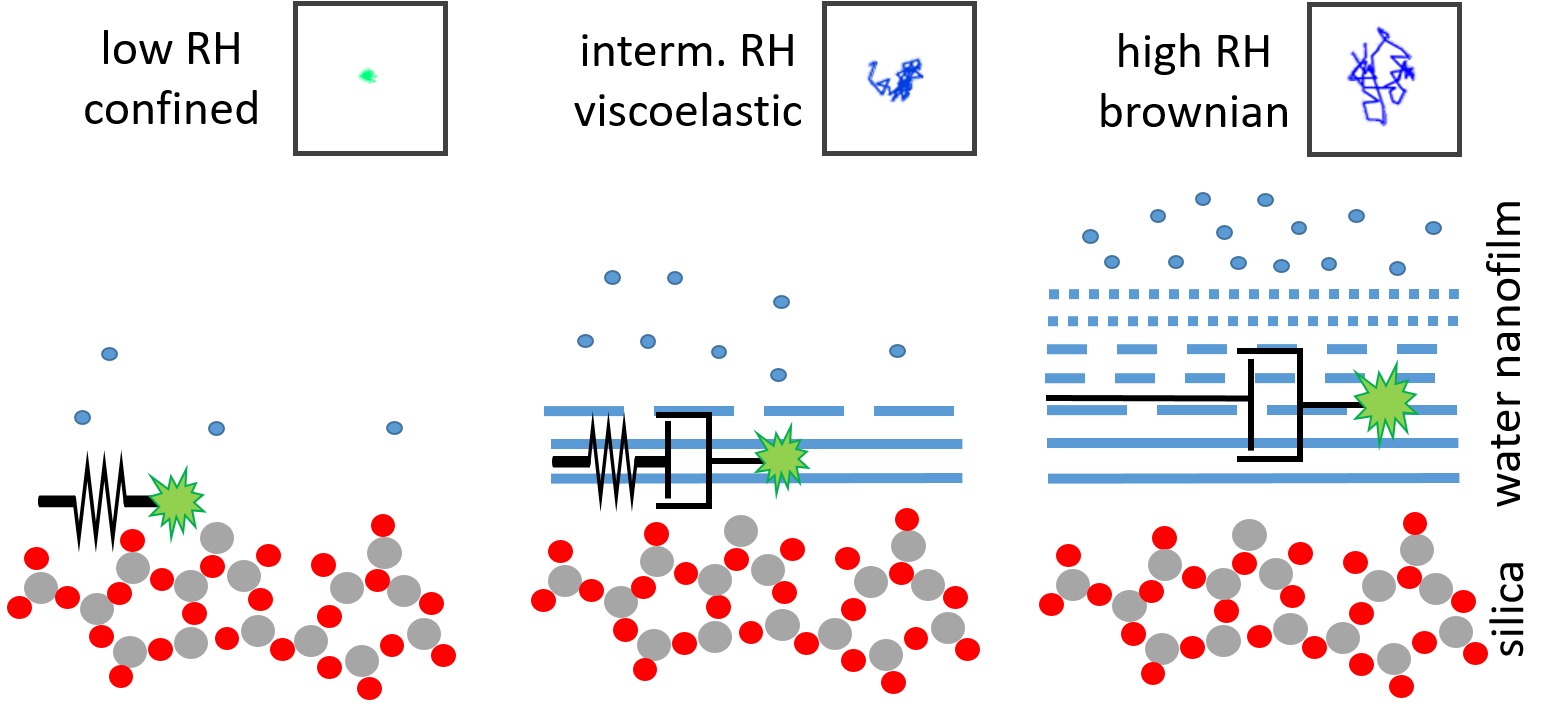}

\end{tocentry}

\begin{abstract}
We used single-molecule tracking to probe the local rheology of interfacial water.
Fluorescent rhodamine molecules were tracked on silica surfaces as a function of ambient relative humidity, which controlled the thickness of condensed water nanofilms. 
At low humidity, the molecules exhibited confined diffusion in the vicinity of isolated adsorption sites characterized by a broad distribution of binding stiffness constants; subsequent chemical or physical surface passivation selectively eliminated stiffer binding sites.
At increased humidity, molecularly thin water films condensed, permitting near-surface transport of rhodamine molecules. 
Motion was subdiffusive, with an anomalous exponent increasing with the nanofilm thickness.
Molecular trajectories were temporally anticorrelated, ergodic, but also featured transient binding and intermittent diffusion.
Statistical modeling demonstrated that this complex motion in water nanofilms had the characteristics of fractional Brownian motion combined with a continuous time random walk. This was consistent with diffusion within viscoelastic nanofilms, suggesting persistent molecular structuring in the vicinity of the silica surface.
\end{abstract}

\section{}
The interface between water and silica (or related silicate minerals) is central to a wide range of geological and industrial processes \cite{Liittge2008,Zhuravlev2000}, yet many of its properties remain elusive, notably in the nanoscale limit where molecules are highly confined and constrained.
The different forms of Si-O groups at the surface are believed to interact strongly with water molecules through hydrogen bonding, leading to molecular ordering that extends several \AA~from the surface \cite{Asay2005}. 
Water structuring, in turn, may impact transport properties at the interface, by influencing the hydrodynamic slip or no-slip boundary conditions \cite{Ortiz2013}, and changing local rheology.
The viscosity of water nanofilms remains hotly debated. 
While studies of water confined between two mica (an aluminosilicate mineral) surfaces separated by 5~nm and less found no substantial increase in viscosity \cite{Israelachvili1986,Raviv2004}, experiments using scanning-probe techniques on silica and other hydrophilic surfaces have reported viscosity enhancement by as much as $10^6$ \cite{Goertz2007,Ortiz2013}.

Hydrophilic surfaces exposed to water vapor become spontaneously coated by a thin film of water at the \AA-to-nm scale, with the exact thickness depending on substrate chemistry, temperature, and vapor pressure. 
Water physisorption is therefore ubiquitous in the Earth's subsurface, surface, and atmosphere, including on aerosol particles and ice (glacier and snow) \cite{Ewing2006}, with significant consequences for environmental processes, for example in heterogeneous chemical reactions \cite{Sumner2004}.
Thus, there is significant interest in understanding molecular transport in water nanofilms, despite the uncertainty with respect to the structure of these thin films \cite{Li2015}. Recent studies have elucidated the mechanisms of molecular interfacial transport through bulk-mediated diffusion, whereby molecules transiently bind to the surface and diffuse mostly in the bulk \cite{Skaug2013,Wang2017}, describing a ``hopping'' diffusion which fits in the framework of continuous-time random walks (CTRW).
However, little is known about how extreme confinement, between two solids or between solid and vapor, perturbs this behavior \cite{Verdaguer2006}. 
Presumably, confinement would increase the amount of interface available and modify the hydrodynamic properties of the fluid phase, which could result in different types of stochastic dynamics.

Here, we report studies of molecular diffusion in water nanofilms, and discuss the implications on the rheology and structure of the aqueous phase. 
Silica surfaces were exposed to environments with controlled relative humidity (RH), resulting in the condensation of water nanofilms, increasing in thickness with RH. 
Single-molecule tracking (SMT) was used to investigate the motion of fluorescent rhodamine molecules in these films. 
These experiments revealed interesting features relative to the energy landscape of the silica surface, and the nature of the near-surface water layer. 
We found that the silica adsorption sites exhibited an unexpectedly wide distribution of restoring forces, and that chemical or physical passivation influenced this distribution in distinctive ways.
In the absence of condensed water, molecules exhibited confined diffusion only in the vicinity of these sites. 
At higher RH, a continuous water nanofilm formed, which enabled diffusive transport across the interface.
Diffusion in the nm-thick water layer was subdiffusive and ergodic, consistent with a viscoelatic-type of diffusion.
This provided important insights into the properties of the first few molecular layers of water adsorbed on silica, and suggested the presence of persistent molecular structuring. 
We employed superstatistical and subordinating models to infer the details of the microscopic processes from the ensemble behavior observed.

\section{Results and Discussion}

\subsection{Evolution of Molecular Mobility with Nanofilm Thickness}

\begin{figure}
\includegraphics{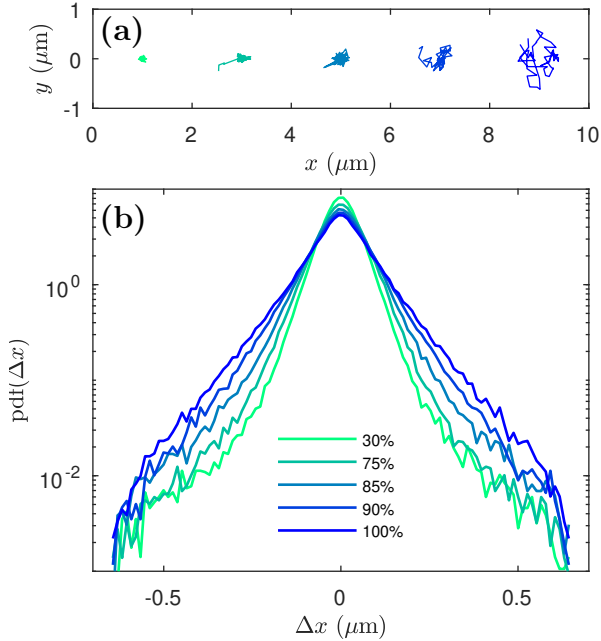}
\setlength{\unitlength}{1cm}
\put(-7.3,8) {\textbf{(a)}}
\put(-7.3,5.5) {\textbf{(b)}}
\caption{\label{fig1}(a) Representative molecular trajectories for RH = (30, 75, 85, 90, 100)\%, from green to blue (left to right), respectively. 
Trajectories evolved from confined to exploratory as RH increased.
(b) One-dimensional displacement distributions at $\Delta t$ = 50~ms for the same RH values.
As RH increased, larger displacements became more likely, and the tails of the distributions became exponential.}
\end{figure}

As RH increased, water nanofilms grew in thickness, and rhodamine molecules became significantly more mobile (see Supplemental Movies), as previously observed \cite{Mitani2006,Giri2015}. 
Specifically, molecular trajectories evolved from confined to exploratory with increasing RH, with exploratory trajectories often exhibiting alternation between slow diffusion and long jumps (Fig.~\ref{fig1}a). 
These intermittent dynamics were qualitatively similar to trajectories observed in SMT experiments at the interface between solid and bulk liquid \cite{Skaug2013,Wang2017}. However, as shown below, nanofilm confinement resulted in new modes of diffusion. 
A more quantitative characterization of this increased mobility was obtained using probability distribution functions (pdf). 
The pdf of one-dimensional displacements $\Delta x = X(t+\Delta t) - X(t)$ of measured positions~$X$ between successive frames ($\Delta t = 50$~ms) were distinctly non-Gaussian, and showed a clear evolution towards a Laplace-like distribution (exponential tails) at high humidity, as seen in  Fig.~\ref{fig1}b.
The evolution of these tails in the distribution quantitatively demonstrated that the increased mobility with nanofilm thickness was due to the increased presence of large displacements, \textit{i.e.} ``flights''. 
Below, the molecular motion under various RH conditions is described in detail.

\subsection{Confined Diffusion on Dry Silica}
As a baseline, we first studied the interaction between the molecular probes and the silica surface in the absence of condensed water.
At 30~\% RH, a sub-monolayer quantity of water was present on the silica surface \cite{Sumner2004}, and diffusion of adsorbed rhodamine probes was almost entirely confined.
The displacement distribution of rhodamine did not change significantly as a function of time interval, and the mean square displacement MSD~$= \langle \left( X(t+\Delta t) - X(t) \right)^2 \rangle$ as a function of lag-time $\Delta t$ was flat (see Supporting Information). 
These metrics are characteristic of confined dynamics, and implied that molecules diffused only in the local vicinity of specific adsorption sites, without escaping small regions defined by the balance between thermal energy $k_BT$ and a site~$i$'s ``stiffness'', $k_i$.
Molecules therefore served as local probes of the surface's energy landscape near local minima.

\begin{figure}
\includegraphics{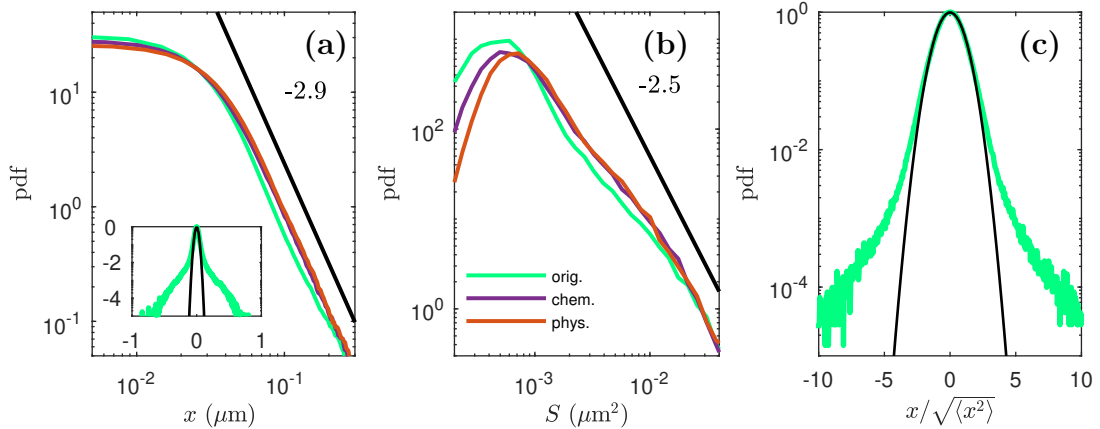}
\setlength{\unitlength}{1cm}
\put(-12,5) {\textbf{(a)}}
\put(-7.5,5) {\textbf{(b)}}
\put(-2.5,5) {\textbf{(c)}}
\caption{\label{fig2}
(a) Distribution of re-centered positions $x$ at RH~=~30\% in a log-log plot, showing a power-law tail (slope: $-\mu$). 
Black line of slope -2.9 added for reference.
Inset: corresponding distribution in log-linear format (log(pdf) vs $x$ in $\mu$m), and quadratic fit (solid black) showing a strong discrepancy with a single harmonic potential hypothesis.
(b) Distribution of compliances $S = k_BT/k$ showing a power-law tail (slope: $-q$). Black line of slope -2.5 is added as a reference.
Green is original (untreated) silica surface, purple after chemical passivation, and orange after physical passivation (see text).
(c) Distribution of positions renormalized by the compliance of their corresponding adsorption site, and quadratic fit. 
The harmonic approximation is adequate to about 4$\sigma$.
Exposure time was 20~ms.
}
\end{figure}

To characterize these adsorption sites, we looked at the distribution $P(x)$ of probe positions $x$ relative to their respective confining sites.
We calculated the re-centered position $x = \lbrace x^{(i)}_t\rbrace$ of trajectory~$i$ at time point~$t$ by selecting long trajectories 
(longer than 20 frames; using a different cutoff of 30 or 50 frames gave indistinguishable results) 
and subtracting from each measured position $X^{(i)}_t$ the mean position of its trajectory: $x_t^{(i)} = X_t^{(i)} - \langle X^{(i)}\rangle_t$.
In the harmonic potential approximation, if all the adsorption sites were equivalent with stiffness $k$, $P(x)$ would be Gaussian: $P(x) \sim \exp \left( -kx^2/2k_BT \right)$ \cite{Lindner2013}. 
Instead, Fig.~\ref{fig2}a shows that $P(x)$ exhibited a power-law dependence at large $x$, $P(x) \sim x^{-\mu}$ with  $\mu \simeq 2.9$. 
Moreover, calculating the MSD for each trajectory revealed a wide distribution of site stiffnesses, with trajectories confined more or less tightly around different adsorption sites (see Supporting Information).
The majority of true molecular motion obviously occurred on softer sites, and in fact, the apparent motion on very strong binding sites (high stiffness, small MSD) was actually due to localization error (about 50~nm\cite{Skaug2013}).
Therefore, focusing on the stiffness distribution of softer sites, we calculated so-called mechanical compliance~$S_i$ \cite{Wang2004}, defined as inverse stiffness, renormalized by $k_BT$ for a direct correspondence with trajectory statistics:  $S_i = k_B T/k_i = \text{MSD}(x^{(i)})$.
Fig.~\ref{fig2}b shows that the compliance distribution~$\Pi(S_i)$ was broad, with a power-law tail of exponent $q \simeq 2.5$. 
This demonstrated that adsorption sites were intrinsically heterogeneous, providing compelling insights into the heavy-tailed distribution of waiting times observed in bulk-mediated diffusion \cite{Skaug2013} and at high RH (see below).

We hypothesized that the presence of heterogeneous binding sites was related to chemical and/or physical surface features of the silica. 
To test this, we modified surfaces to reduce the density and magnitude of anomalously strong binding sites. 
Specifically, we gently passivated silica surfaces by chemical or physical methods, using either adsorption of polyethylene glycol (PEG) at low concentration (chemical passivation), or thermal annealing (physical passivation), see Methods.
Adsorption of PEG at very low coverage has been shown to selectively block anomalously strong adsorption sites \cite{Alcantar2000,Morrin2018}, and thermal annealing is known to smooth \AA-scale topographical features \cite{Persson2019}.
As shown on Fig.~\ref{fig2}b, the chemically and physically passivated surfaces exhibited a reduction in strong binding sites by factors of $\simeq3$ and $\simeq10$, respectively, while the power-law tails associated with weak binding sites remained essentially unchanged.
These observations confirmed the origins of the distribution of site compliances as a combination of molecular-scale chemical and physical heterogeneities.
In addition, while PEG and rhodamine are expected to have different binding affinities to adsorption sites, we observed that after PEG passivation the distribution of adsorption spring constants for rhodamine was modified only for high-stiffness sites.
This suggests that adsorption sites were non-specific, and likely similar for PEG and rhodamine.

Despite surface passivation, which quenched only the highest-energy binding sites, the overall confined dynamics, and notably the power-law tails of probe positions, remained unaffected, consistent with expectations that the majority of molecular motion occurred on softer sites (Fig.~\ref{fig2}a).
Indeed, one can show mathematically that the heavy tail of compliances produces the heavy tail of  $P(x)$, using a simple superstatistical model, whereby the distribution for a unique site is averaged over the distribution of all site compliances \cite{Chechkin2017}. 
Assuming a harmonic potential approximation for each adsorption site and a compliance distribution $\Pi(S) \sim S^p/(S_0 + S)^{p+q}$, the distribution $P(x)$ is given by
\begin{equation}
P(x) \sim  \int \frac{S^p}{(S_0+S)^{p+q}} \cdot e^{-x^2/2S} \cdot dS \sim U(q-1,-p,x^2/2S_0),
\end{equation}
where $U$ is the Tricomi function. 
Notably, $U(q-1,-p,x^2/2S_0) \sim x^{-(2q-2)}$ at large $x$, and we recover the tail exponent of $P(x)$: $\mu = 2.9 \simeq 2q-2$ when $q = 2.5$. 
In addition, if positions $x^{(i)}_t$ are renormalized by the standard deviation of their respective trajectories, $\sigma_i = \sqrt{S_i}$, the resulting distribution is better approximated by a Gaussian (see Fig.~\ref{fig2}c).
These results show that binding site heterogeneity broadens ensemble statistics, and that blocking of anomalously strong binding sites does not critically alter the observed dynamics.
In addition to trapping rhodamine molecules more tightly, it is also possible that these strong binding sites acted as nucleation points for the condensation of water vapor into liquid nanofilms at higher values of relative humidity, due to their non-specificity.

\subsection{Non-Brownian Diffusive Transport at High Humidity}
\begin{figure}
\includegraphics{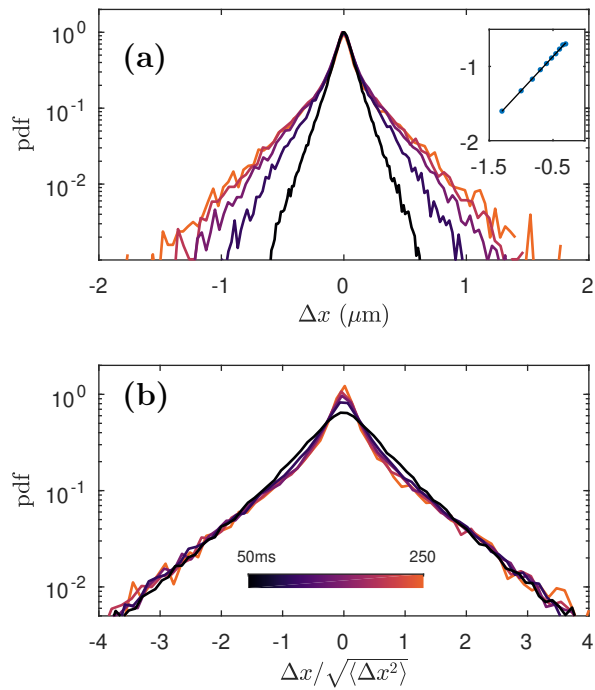}
\setlength{\unitlength}{1cm}
\put(-7,8.5) {\textbf{(a)}}
\put(-7,4) {\textbf{(b)}}
\caption{\label{fig3}
(a) Probability distribution functions of displacements $\Delta x$ at different time-lags $\Delta t$ (50 to 250~ms; purple to orange respectively) for RH = 100\%. Inset: log(MSD) vs log($\Delta t$), and linear fit giving a slope of 0.92. 
(b) Pdf of displacements rescaled by their corresponding standard deviations, showing self-similar exponential tails.
}
\end{figure}

When water condenses on the silica surface, water molecules compete with probe molecules for adsorption sites, and provide a medium for interfacial transport.
Indeed, at 100\% RH, the surface was covered by a water film averaging 8 water layers \cite{Sumner2004}, and molecular trajectories were not confined, revealing that the presence of condensed water enabled long-range diffusive transport across the surface. 
The pdf of displacements at different lag-times $\Delta t = n t_E$ (with $t_E = 50$~ms the exposure time) showed a clear broadening with increasing $\Delta t$, and the corresponding MSD scaled almost linearly with time: MSD $\sim \Delta t^\alpha$ with $\alpha = 0.92$ (Fig.~\ref{fig3}a).
However, the displacement distributions were distinctly non-Gaussian, indicating that probe molecules did not undergo regular Brownian motion.
Furthermore, after rescaling by their corresponding standard deviations $\sigma_n = \sqrt{\langle \left( X(t+nt_E) - X(t) \right)^2 \rangle}$, these distributions exhibited self-similar exponential tails, and did not tend asymptotically towards a Gaussian distribution at measurable timescales (Fig.~\ref{fig3}b).
At longer lag-times, a central peak emerged.
Combined, these observations suggested a mechanism fundamentally different from ``anomalous-yet-Gaussian" diffusion \cite{Wang2009}, and the presence of the central peak hinted at intermittent diffusion, in which transient periods of ``crawling'' diffusion of surface-bound molecules alternate with long ``flights'' in the water nanofilm.
The crawling periods lead to a high probability of small displacements, \textit{i.e.} the central peak.
The possible origins of exponential tails are diverse, but Laplace distributions can result from the superposition of Gaussian processes with different parameters, such as an exponentially distributed diffusion coefficient \cite{Lampo2017}.
It is therefore plausible that the exponential tails arise from a combination of different modes of diffusion within the water nanofilm. 
In order to better understand these processes, we examined the diffusive behavior at intermediate values of RH.


\subsection{Subdiffusive Exponent Evolution with Nanofilm Thickness}
\begin{figure}[]
\includegraphics{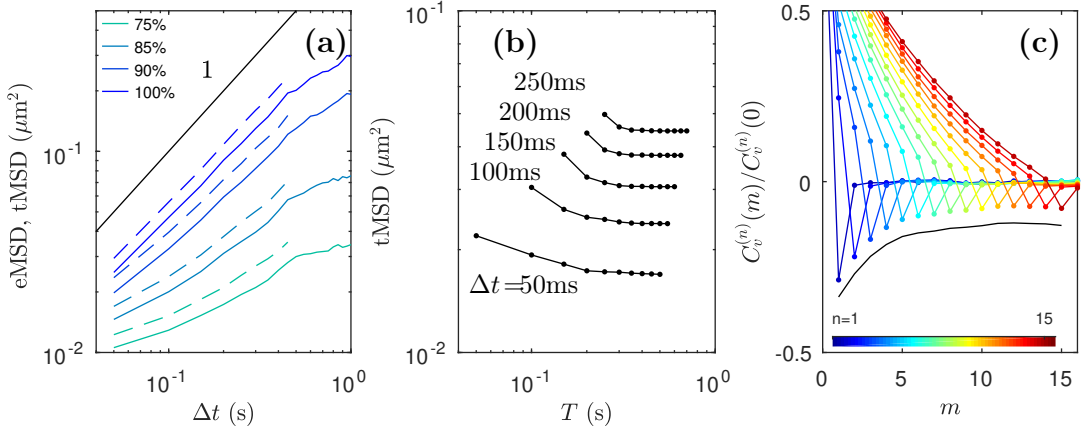}
\setlength{\unitlength}{1cm}
\put(-12,5) {\textbf{(a)}}
\put(-9.5,5) {\textbf{(b)}}
\put(-2.5,5) {\textbf{(c)}}
\put(-9.8,1.8) {\footnotesize{$\Delta t \! = \! $50ms}}
\put(-9.8,3.3) {\footnotesize{100ms}}
\put(-9.6,3.7) {\footnotesize{150ms}}
\put(-9.4,4.1) {\footnotesize{200ms}}
\put(-9.2,4.5) {\footnotesize{250ms}}
\caption{\label{fig4}
(a) eMSD (solid) and tMSD (dashed) curves as a function of lag-time for RH = (75, 85, 90, 100)\%, from green to blue. Black line of exponent 1 added for reference.
(b) tMSD curves (RH = 85\%) as a function of experimental time $T$, for five consecutive lag-times $\Delta t = 50 \dots 250$~ms.
(c) Velocity auto-correlation curves (RH = 85\%) for $n$ = 1 to 15 (blue to red) showing a persistent negative peak at $n=m$. Black curve is a guide to the eye showing the relative position of the peaks.
}
\end{figure}

For RH $\geq$ 75\%, multiple layers of water were present, and molecular trajectories evolved from subdiffusive to nearly Brownian with increasing RH, as indicated by the time-dependence of the ensemble MSD (eMSD) in Fig.~\ref{fig4}a.
In particular, power-law fits (least-square of logarithms) to the eMSD~$\sim \Delta t^\alpha$ yielded values: 
$\alpha = 0.44, 0.62, 0.82, 0.93$ for RH = (75, 85, 90, 100)\%, respectively, approaching unity at high humidity.
Since the origins of subdiffusion may be diverse, a deeper characterization of the dynamics was necessary to better identify the underlying mechanisms.


First, we calculated the time-averaged MSD for each single-molecule trajectory, using the conventional definition \cite{Metzler2014,Tabei2013,Klafter2011}, 
\begin{equation}
\overline{\delta^2(\Delta t,T)} = \frac{1}{T{-}\Delta t} \int_0^{T-\Delta t} \left[ X(\Delta t + t'){-}X(t')\right]^2 dt',
\end{equation}
and then considered the average over the ensemble of trajectories: 
tMSD = $\langle \overline{\delta^2(\Delta t,T)} \rangle_\mathrm{ens}$.
The behavior of the tMSD, as a function of both lag-time $\Delta t$ and experimental time $T$, provides information about the presence of aging effects (non-ergodicity) in the observed dynamics.
We found that the lag-time dependence of the tMSD displayed very similar slopes as the eMSD counterparts, as shown on Fig.~\ref{fig4}a (fit values $\alpha = 0.49, 0.66, 0.84, 0.95$, within 0.05 of eMSD fits).
Moreover, the tMSD was independent of experimental time $T$, except at short~$T$ (Fig. \ref{fig4}b and Supporting Information).
Combined, these observations indicated that the diffusive process was ergodic\cite{Metzler2014}.

A second important result was that trajectories exhibited temporally anticorrelated displacements. 
Reliably measuring velocity correlations in SMT experiments is difficult due to the high level of localization error (static and dynamic)), which leads to apparent anticorrelation even in the absence of truly anticorrelated motion. 
However, an analytical method was recently developed to circumvent these difficulties and extract the signature of anticorrelated motion from noisy localization data \cite{Weber2012}.
We calculated velocity autocorrelation (VAC) functions, $C_v^{(n)}(m)$, at different lag-times $\Delta_m = mt_E$ and for average velocities over time intervals $\Delta_n = nt_E$ according to:
\begin{equation}
C_v^{(n)}(m) = 
 \left< \frac{\overrightarrow{r}(t{+}\Delta_m{+}\Delta_n) {-} \overrightarrow{r}(t{+}\Delta_m)}{\Delta_n} \cdot  \frac{\overrightarrow{r}(t{+}\Delta_n) {-} \overrightarrow{r}(t)}{\Delta_n}  \right>_t.
\end{equation}
As shown in Fig.~\ref{fig4}c, the VAC functions exhibited a persistent negative peak when $m=n$, which is a robust signature of temporally anticorrelated dynamics, even in the presence of moderate localization error \cite{Weber2012}.
In particular, the negative asymptotic value at large $n$ demonstrates a true anticorrelation \cite{Weber2012,Backlund2015}, which is present at all values of RH but decreases in magnitude at high RH (see Supporting Information).

Taken together, these observations demonstrated that transport in water nanofilms was subdiffusive, ergodic, and temporally anticorrelated, which in turn suggested a mechanism described by fractional Brownian motion (fBM) \cite{Metzler2014}.
However, the full picture is slightly more complex, due to the likelihood of transient binding with the silica surface.
In the next section, we propose a model of anomalous diffusion coupled to transient binding which reflects the observed statistics.


\subsection{Discussion}

\begin{figure}[]
\includegraphics{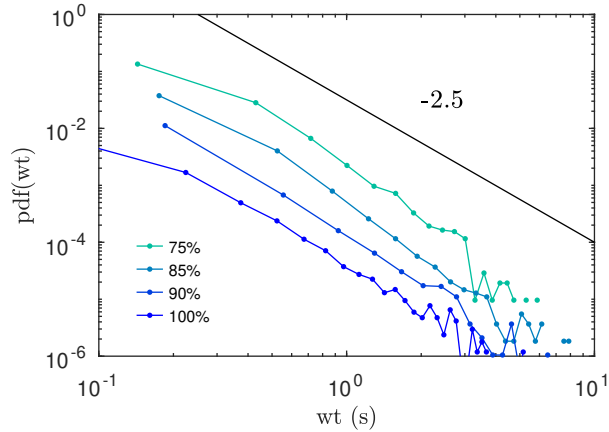}
\caption{\label{fig5}
Waiting times (wt) distributions for RH = (75, 85, 90, 100)\% (green to blue), with artificial offset for clarity. Black line of exponent -2.5 added for reference.
}
\end{figure}

Due to short-range attractive interactions between rhodamine molecules and the silica surface, one expects transient binding, resulting in intermittent diffusion, usually described in the framework of continuous time random walks (CTRW). 
A CTRW process involves alternating waiting times and displacements, where the waiting time intervals and the displacement distances are not constant, but drawn from distributions, which may be heavy-tailed (\textit{e.g.} power-law). 
This has been demonstrated for interfaces with bulk water \cite{Skaug2013,Wang2017}.
The CTRW model is especially compelling here because it has been proven to be the underlying mechanism in bulk-mediated diffusion, and, as described above, rhodamine  molecules adsorb at specific sites with various strengths on dry silica surfaces. 
To characterize the time intervals associated with transient adsorption in nanofilms, we measured the distribution of waiting times between displacements larger than a threshold set at 0.1~$\mu$m, which reflects near-immobility. 
The conclusions were insensitive to the specific choice of threshold (see Supporting Information). 
Fig.~\ref{fig5} shows that the waiting time distribution $\psi(t)$ did not depend on RH, and was indeed heavy-tailed: $\psi(t) \sim 1/t^{1+\beta}$, for large $t$, with $\beta \simeq 1.5$. 
(This exponent is similar to previously observed values \cite{Skaug2013}.)
These distributions demonstrate the presence of transient binding, and the power-law tail is indicative of substantial surface heterogeneities, which is consistent with the observed confined diffusion on dry silica.

The CTRW mechanism alone, however, cannot explain the overall observed subdiffusion, because CTRWs with the waiting time distribution described above ($\beta = 1.5$)is not subdiffusive \cite{Metzler2000}. 
However, we show below that CTRW and fBM may co-exist, by describing a simple model of fBM subordinated to a CTRW which is consistent with the experimentally observed statistics.
Practically, this means that the number of fBM ``steps'' taken per unit time is determined by the embedding CTRW process.
We follow the reasoning of Meroz \textit{et al.} \cite{Meroz2010}, but adapt it to reflect our experimental observations. 
The key difference is that the distribution of waiting times observed here has an asymptotic exponent of $1+\beta = 2.5 > 2$, and therefore possesses a first moment $\overline{\tau}$. 
This has implications on the mathematical derivation, and its outcome; fBM surbordination to CTRW with $\beta < 1$ has been previously investigated, and results in a process with weak ergodicity breaking \cite{Tabei2013}.

We consider a CTRW with a distribution of waiting times $\psi(t)$ for which the walking phases are not Brownian, but fBM in nature.
We propose to calculate the tMSD for this mixed model, $\langle \langle \delta^2 (\Delta t) \rangle_{T} \rangle_\mathrm{ens}$ for lag-times $\Delta t$ within a given experimental time $T$, and averaged over an ensemble of trajectories.
The derivation is detailed in the Methods section.
We obtain tMSD $\sim \Delta t^\alpha$, with same exponent as the eMSD and no dependence on $T$.
This model therefore predicts a tMSD that has the same dependence on $\Delta t$ and $T$ as observed in the reported tracking of rhodamine molecules in Fig.~\ref{fig4}b, and explains the presence of temporal anticorrelation.

Although it provides a clear picture of a possible mechanism for the observed dynamics, there could also be alternative or complementary mechanisms at play.
For example, local confinement around heterogeneous binding sites could also induce a persistent anticorrelation in measured trajectories due to reflections within the trap; however, the apparent ergodicity of the trajectories points rather to motion through a medium with slow stress relaxation, \textit{i.e.} viscoelastic.

\section{Conclusion}

While single-molecule tracking is frequently used to study the dynamics of molecules within a specific environment in an effort to understand the mechanisms of mass transport, here we employ molecular probes to infer information about the local rheology of water nanofilms on silica.
This approach complements microscopic measurements of interfacial water structure (from molecular dynamics simulations or scattering measurements) and force measurements (from  colloidal force microscope or surface force apparatus experiments).
Thus, instead of inferring mass transport dynamics from structural information, we directly measured molecular diffusion in a constrained solvent.
Importantly, these experiments provide direct evidence that probe molecules within nanofilms on silica became continuously more mobile as the thickness of the condensed water nanofilms increased.
Transport was markedly subdiffusive within nanofilms averaging a few molecular layers in thickness, with features typical of fBM. 
Fractional Brownian Motion describes a random process in which successive steps are not independent, but rather depend on the history of the trajectory.
For fBM specifically, the memory kernel, i.e. the extent of correlations between two time points, decays as a power law of the time difference.
The anticorrelation observed between displacements signified that when a displacement occured in one direction, the following displacement was more likely to occur in the opposite direction rather than the initial one.
This suggests an effective restoring force, and hence is reminiscent of diffusion in viscoelastic media, as previously observed in crowded or gel-like environments, such as in the cellular cytoplasm \cite{Backlund2015}. 
It is surprising to observe the same dynamics in a system as simple as a water film adsorbed on silica, and it is presumably indicative of molecular structuring within the water phase, and its consequential departure from purely viscous behavior.
Of significant practical interest was the fact that molecular mobility increased continuously with the nanofilm thickness.

This was first observed in Ref.~\cite{Mitani2006}, where macroscopic measurements revealed an increase of the diffusion coefficient by almost two orders of magnitude between intermediate (RH = 80\%) and high (RH = 95\%) humidity.
Our SMT experiments demonstrated that increasing humidity alters not just effective diffusion coefficients, but most notably the subdiffusive exponent $\alpha$, making relative humidity a single parameter that fundamentally changes the properties of the water thin film.
This finding could find numerous applications in nanotechnologies, whereby the dynamic adjustment of humidity conditions would drastically affect molecular transport on a solid surface, from fully confined to diffusive, and hence allow or restrict search processes or reactivity.
Finally, besides the results pertaining to the silica-water interface, the method described in this paper can be easily extended to other liquid-solid interfaces, and molecular probes can be intentionally designed to investigate specific properties of near-surface liquid films.

\section{Methods}

\subsection{Experiments}

Borosilicate glass coverslips (Fisher Scientific) were cleaned by immersion in piranha solution (70\% sulfuric acid, 30\% hydrogen peroxide solution) for 1~h, followed by exposure to UV/ozone for 30~min.
Then, 6~$\mu$L of $10^{-8}$~M rhodamine 6G (Sigma) in a 50\% solution of methanol in deionized water were deposited on the coverslips, and dried in a vacuum chamber for 30~min. 
The substrates were immediately isolated from the atmosphere and placed in the imaging setup. 
They exhibited total wetting for well over one hour after preparation. 
Total Internal Reflection Fluorescence (TIRF) microscopy was performed using a Nikon Eclipse TI93 microscope with a 100x objective and an EMCCD camera (Photometrics Cascade 512B) for image acquisition. 
A 532~nm laser (Cobolt Samba) was used for excitation, and the acquisition time was $t_E$ = 20~ms or 50~ms. 
Custom MATLAB code was employed for particle tracking and data analysis. 
The amount of adsorbed water was tuned by changing the relative humidity (RH) in the vicinity of the coverslip, in the range of 30\% RH to 100\% RH, which corresponded to an equivalent number of water layers between $\simeq$1 and 8 \cite{Sumner2004}. 
The relative humidity in equilibrium with the coverslip was determined by creating a small, sealed volume using CoverWell chambers (Grace Bio-Labs) in equilibrium with saturated water solutions of different salts. Following Ref.~\cite{Rockland1960} 
, we used saturated solutions of Magnesium Chloride (Sigma), Sodium Chloride (Fisher), Potassium Chloride (Sigma), and Barium Chloride (Sigma), which have equilibrium RH at 25$^{\circ}$C of 32\%, 75\%, 85\%, 90\%, respectively and with uncertainty within $\pm$1\%.
A RH of 100\% was achieved using deionized water instead.
The vapor exchange between the sealed volume and the solutions was made through small holes at the top of the CoverWell chamber, which were directly in contact with solutions droplets, for at least 10~min.

Surface modification for experiments at RH = 30\% was performed as follows.
For chemical passivation, 3~$\mu$L of polyethylene glycol (molecular weight 10,000 g/mol, Sigma) at a concentration of $10^{-4}$~M in a 50\% methanol solution were deposited on a clean coverslip and let dry before depositing rhodamine. 
Assuming a radius of gyration of about 5~nm~\cite{Linegar2010} and given the size of the coverslip, 25$\times$25~mm$^2$, this corresponds to a surface coverage of less than 0.1\% (total wetting was achieved).
For physical passivation, coverslips were left in a furnace at 500~C for about 100~h before the cleaning and deposition procedures.

For each surface and humidity condition, we acquired between 10,000 and 20,000 single-molecule trajectories, corresponding to an order of 10$^5$ individual displacements from which our statistics were derived.

\subsection{Mathematical Derivation}

We consider a CTRW with a distribution of waiting times $\psi(t)$ for which the walking phases are not Brownian, but fBM in nature.
We propose to calculate the tMSD for this mixed model, $\langle \langle \delta^2 (\Delta t) \rangle_{T} \rangle_\mathrm{ens}$ for lag-times $\Delta t$ within a given experimental time $T$, and averaged over an ensemble of trajectories.
We first note that, by commutation of the averaging procedure:
$\langle \langle \delta^2 (\Delta t) \rangle_{T} \rangle_\mathrm{ens} = \langle \langle \delta^2 (\Delta t) \rangle_\mathrm{ens} \rangle_{T}$.
Next, we introduce the number of steps $n(\tau)$ of the random walk in a time interval $\tau$, and from Ref.~\cite{Meroz2010} we have $\langle \langle \delta^2 (\Delta t) \rangle_\mathrm{ens} \rangle_{T} \sim \langle \langle n^\alpha(\Delta t) \rangle_\mathrm{ens} \rangle_{T}$ . 
The calculation of the right-hand side requires the time $T$-integration of the ensemble average
\begin{equation}
\langle (n(t{+}\Delta t) {-} n(t))^\alpha \rangle_\mathrm{ens} {=} \! \!
\int_0^{\Delta t} \! \! \! \! \langle n^\alpha(\Delta t {-} t_w) \rangle_\mathrm{ens} \psi_1(t_w|t) dt_w,
\end{equation} 
where $\psi_1(t_w|t)$ is the pdf of the forward waiting time, \textit{i.e.} the time $t_w$ between $t$ and the next following step \cite{Meroz2010,Klafter2011}. 
Because $\psi$ possesses a first moment, we have: 1) $\langle n^\alpha(t) \rangle \sim t^\alpha$ \cite{Blumen1984}; and 2) $\psi_1(t_w|t)$ quickly reaches an equilibrium form which  depends only on $t_w$ and not on $t$: $\psi_1 \simeq \psi_1^{eq} \sim 1/\overline{\tau} t_w^\beta$, according to Ref.~\cite{Klafter2011}. 
Treating the part of the integral near 0 as a constant, in the limit of large $\Delta t$ we get
\begin{equation}
\langle (n(t+\Delta t) - n(t))^\alpha \rangle_\mathrm{ens} \sim \int_\epsilon^{\Delta t} (\Delta t -t_w)^\alpha t_w^{-\beta} dt_w \sim \Delta t^\alpha.
\end{equation}
(See Supporting Information for more details.) 
After $T$-integration, we obtain tMSD $\sim \Delta t^\alpha$, with same exponent as the eMSD and no dependence on $T$.
\begin{acknowledgement}


The authors thank D. Kienle for providing some of the particle tracking code.
This work was supported by the U.S. Department of Energy, Office of Science, Basic Energy
Sciences, under Award \#DE-SC0001854.

This article is dedicated to Professor Peter S. Pershan, a pioneer in experiments that probed the near-surface structure of liquids and liquid crystals, on the occasion of his 85th birthday.

\end{acknowledgement}

\begin{suppinfo}

%
The Supporting Information is available free of charge at https://pubs.acs.org/doi/10.1021/
\begin{itemize}
  \item supporting\_info\_text.pdf: text containing additional experimental results and details on the mathematical derivation
  \item diffusion\_RH=75.avi: movie of molecular diffusion at RH = 75\%
  \item diffusion\_RH=100.avi: movie of molecular diffusion at RH = 100\%
\end{itemize}

\end{suppinfo}

\bibliography{achemsoBIB}

\end{document}